# Pseudospin Hall Transport Induced by Berry Curvature

Qinhui Jiang[1,*], Jidong Song[1,*], Qingyang Mo[2], Bo Li[1,3], Dongyi Wang[2,†], Shuang Zhang[2,4,5,6,7,†], and Mengyao Li[1,†]

[1] Institute of Materials Research, Tsinghua Shenzhen International Graduate School, Tsinghua University, Shenzhen 518055, China

[2] New Cornerstone Science Laboratory, Department of Physics, The University of Hong Kong, Hong Kong 999077, China

[3] Suzhou Laboratory, Suzhou 215000, China

[4] Materials Innovation Institute for Life Sciences and Energy (MILES), HKUSIRI, Shenzhen, China

[5] State Key Laboratory of Optical Quantum Materials, The University of Hong Kong, 999077 Hong Kong, China

[6] Department of Electrical and Electronic Engineering, The University of Hong Kong, 999077 Hong Kong, China

[7] Quantum Science Center of Guangdong-Hong Kong-Macao Great Bay Area, 3 Binlang Road, Shenzhen, China

Correspondence: Dongyi Wang, Shuang Zhang, Mengyao Li

## Abstract

Pseudospin-1 Dirac systems exhibit unique physics distinct from conventional Dirac cones, such as flat-band crossings and non-Abelian characteristics, yet their topological transport properties have remained largely untapped in passive, time-reversal-invariant settings. Here we uncover an in-plane polarity of the Berry-curvature texture in a pseudospin-1 Dirac Hamiltonian, a previously unexplored geometric degree of freedom encoded in the sign-resolved distribution of Berry curvature despite zero net Berry flux, and reveal a new mechanics where Berry curvature induce pseudospin Hall behaviors in a system. We show that the oriented coupling between this momentum-space polarity and a real-space mass gradient governs a geometric selection rule that dictates the emergence of gapless pseudospin Hall modes. By engineering the intracell couplings of a four-site planar lattice, we independently program the Berry-curvature polarity and the spatial mass gradient without altering the host lattice symmetry. Acoustic experiments directly confirm this directional selection rule: reversing the mass gradient closes or reopens the dispersive gap, while pseudospin-selective source excitation launches counterpropagating pseudospin branches along arbitrary prescribed axes. Our work establishes quantum geometric polarity as a versatile tool for reconfigurable wave routing, sensing, and high-capacity quantum applications.

## Introduction

Topological boundary states have reshaped the understanding of wave manipulation across electronic[1], photonic[2,3], acoustic[4,5], and mechanical[6] platforms. The hallmark feature of these states, directionally protected propagation with inherent robustness against disorder, has motivated extensive efforts to engineer boundary waveguides[7–15], localized cavities[16,17], and topological lasers[18–21]. More recently, spatial mass modulation in adiabatic topological waveguides has emerged as a powerful strategy to overcome the constraints of abrupt material interfaces[22–25]. By continuously grading the mass term over multiple unit cells, adiabatic waveguiding reduces radiative leakage in open systems[24], protects modes against sudden symmetry-breaking perturbations, and enables continuous pseudospin switching and phase control[26].

Most implementations of topological waveguiding rely on standard two-band or four-band Dirac and Dirac-like cones, where topological phase transitions are governed by band inversion and quantized non-zero Chern or valley-Chern invariants[27–32]. In contrast, spin-1 and pseudospin-1 Dirac system display different physical characteristics[33–42]. Typically formed by the intersection of a flat band with two linearly dispersing bands, these systems provide fertile platforms for exploring high-degeneracy band crossings[43,44], flat-band localization[45,46], and non-Abelian physics[47]. However, their topological transport properties in passive, reciprocal media remain largely undeveloped. In most canonical planar pseudospin-1 lattices, including Lieb and dice lattices, the net Berry flux around the band crossing yields trivial topological indices, and topologically protected boundary modes are therefore typically absent unless additional mechanisms such as dynamic Floquet engineering[48,49] or active time-modulation[50] are introduced. Consequently, whether pseudospin-1 Dirac systems can host robust, directional topological channels has remained an unresolved question.

In this work, we reveal that a pseudospin-1 Dirac Hamiltonian model with three internal degrees of freedom possesses an oriented, in-plane Berry-curvature polarity that dictates adiabatic boundary transport. In such a system, the local Berry curvature exhibits a sign-resolved, dipolar texture in momentum space, which can induce pseudospin Hall transport. Under spatial confinement, this internal geometric polarity couples directly to the real-space adiabatic mass gradient, establishing a strict geometric selection rule. This matching rule determines whether the transverse confinement opens a trivial gap or cancels the transverse zero-point energy to yield gapless, helical pseudospin Hall modes. We experimentally realize this mechanism in a four-site planar acoustic lattice, where adjustable intracell couplings independently program the internal Berry-curvature polarity and the real-space mass gradient. Both spectral and real-space measurements demonstrate polarity-dependent gap closing and waveguiding along arbitrary in-plane directions regardless of the underlying lattice symmetry. Furthermore, branch-selective helical excitation directly confirms pseudospin-momentum locking along the programmed channels, establishing Berry-curvature polarity as a versatile degree of freedom for topological wave routing.

## Results

### Adiabatic pseudospin Hall response governed by Berry-curvature polarity

We consider a two-dimensional adiabatic pseudospin-1 system (Fig. 1a). Near the Brillouin-zone center Γ, the effective low-energy Hamiltonian reads $H_{2D}(\mathbf{k}; m, \alpha) = k_x S_x + k_y S_y - m(\mathbf{r})\big(Q_{xz}\cos\alpha + Q_{yz}\sin\alpha\big)$, where the mass $m(\mathbf{r})$ couples the three pseudospin components through quadrupolar terms with the internal orientation $\alpha$ (Supplementary Note 1). At zero mass, a flat band intersects two linearly dispersing bands at Γ, forming an isolated pseudospin-1 Dirac crossing (Fig. 1b). Introducing a finite mass lifts threefold degeneracy and opens an energy gap between the dispersive bands while chiral symmetry pins the flat band at zero energy.

For either dispersive band at finite mass, the out-of-plane Berry curvature takes the form $\Omega_z^{(\nu)}(\mathbf{k}; m, \alpha) = \kappa_\nu \frac{m\, \mathbf{p}_\Omega \cdot \mathbf{k}}{\left(k_x^2 + k_y^2 + m^2\right)^2}$, where $\mathbf{p}_\Omega$ denotes the in-plane Berry-curvature polarity vector. The curvature exhibits an odd angular symmetry around Γ, reversing sign across the axis perpendicular to $\mathbf{p}_\Omega$. Although the integrated net Berry flux over the Brillouin zone vanishes, the sign-resolved sectors carry an oriented dipolar polarity. As shown in Fig. 1b, the red and blue regions schematically represent the positive and negative parts of this Berry-curvature texture, respectively, and the purple vector $\mathbf{p}_\Omega$ identifies its global polarity direction. Tuning the internal mass-coupling angle rotates the Berry-curvature polarity continuously in the 2D plane without altering the isotropic dispersion of the host lattice.

The coupling of this polarity to the directed mass texture determines the emergence of the pseudospin bands. For the two bands nearest the flat band, their dispersion is

$$\epsilon_\pm(k_\parallel) = \pm\sqrt{k_\parallel^2 + g(1-\chi)}, \qquad \chi = (\mathbf{p}_\Omega \times \mathbf{g}_m)\cdot\hat{\mathbf{z}}, \tag{1}$$

where $g = |\nabla m| > 0$ is the magnitude of the mass gradient, $\mathbf{g}_m = \nabla m/|\nabla m|$ is the normalized mass-gradient vector, and $k_\parallel$ is the momentum along the isomass line. The term $\chi$ is a geometric factor characterizing the relative orientation between the Berry-curvature polarity and the mass texture. When the polarity and mass-gradient directions are orthogonal with positive orientation, $\chi = +1$, the geometric contribution cancels the zero-point energy of transverse confinement, producing a gapless crossing at Γ that supports helical propagation (Fig. 1d). Reversing the mass evolution yields $\chi = -1$, which removes the cancellation and leaves the two bands gapped (Fig. 1c). Intermediate orientations continuously tune the gap size. This geometric matching condition ($\chi = +1$) provides a deterministic rule for launching adiabatic pseudospin Hall modes by coordinating the internal Bloch-state geometry with the spatial mass profile.

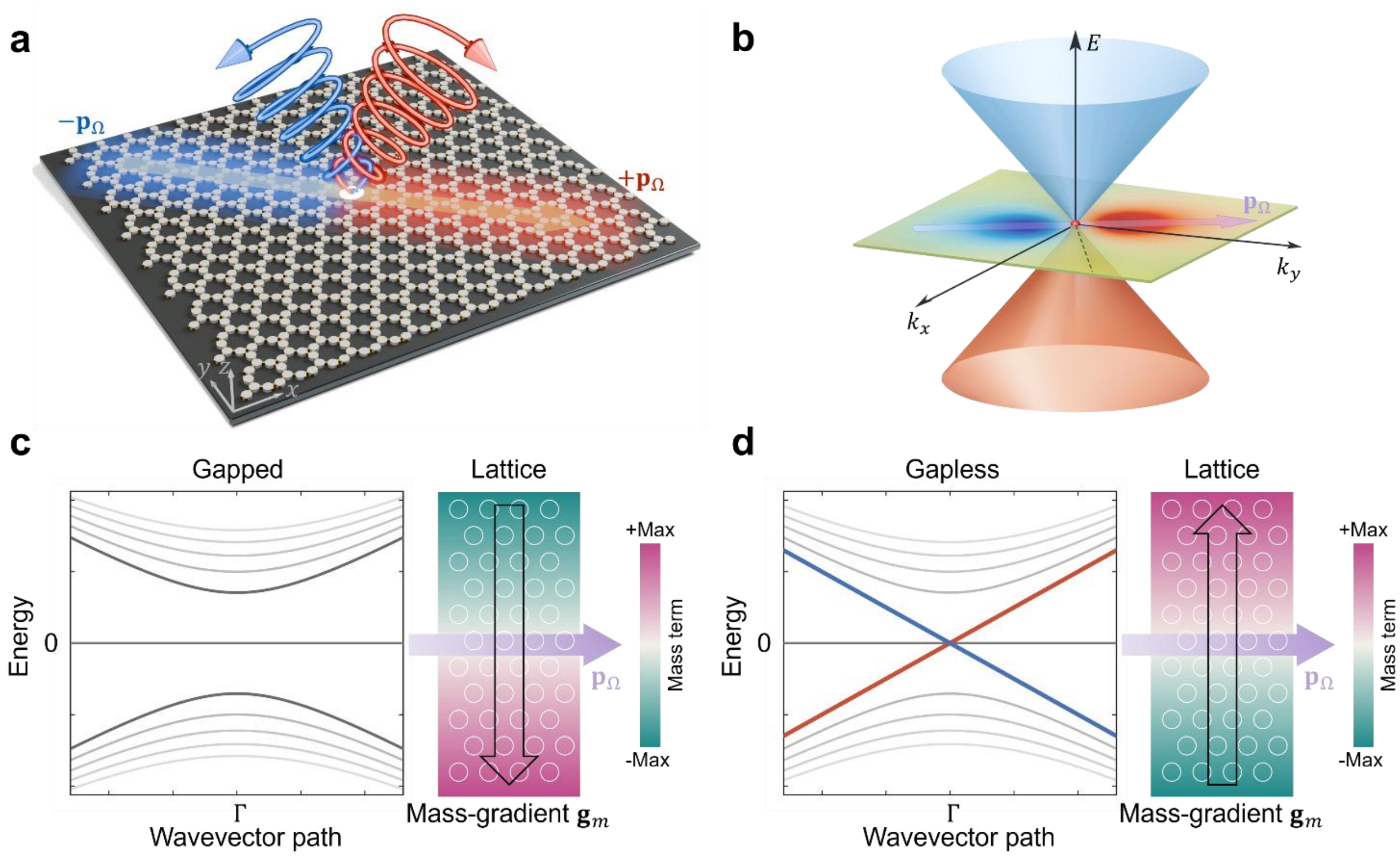

**Fig. 1 Berry-curvature polarity and the adiabatic pseudospin Hall response. a**, Conceptual illustration of a bulk-embedded pseudospin Hall waveguide in a two-dimensional pseudospin-1 lattice. The red and blue helices represent modes of opposite pseudospin propagating along the Berry-curvature polarity $+\mathbf{p}_\Omega$ and $-\mathbf{p}_\Omega$, respectively. **b**, Schematic Γ-centred pseudospin-1 Dirac spectrum,

comprising a flat band and two linearly dispersing bands, together with its odd Berry-curvature polarity texture in the $(k_x, k_y)$ plane. Red and blue regions denote positive and negative Berry curvature, respectively, and the purple vector $\mathbf{p}_\Omega$ indicates its global polarity direction. **c**,**d**, Schematic spectra of modes confined by two opposite linear mass profiles, $m(\mathbf{r})$ and $-m(\mathbf{r})$. In the mismatched configuration (**c**), the two dispersive bands nearest the flat band remain separated by a finite gap at Γ. In the matched configuration (**d**), the counterpropagating pseudospin bands (red and blue) cross at Γ. Pale-grey curves denote bulk bands, the horizontal grey line is the flat band, white circles indicate the lattice unit cells, and the colour bars and black arrows show the signed mass profile and its directed evolution, respectively.

## The planar lattice for programmable Berry-curvature polarity and adiabatic mass textures

To implement this geometric mechanism, the internal Berry-curvature polarity $\mathbf{p}_\Omega$ must be decoupled from the spatial mass-gradient $\mathbf{g}_m$. We achieve this in the four-site planar lattice shown in Fig. 2a. Three outer sites form the propagating network through fixed intercell couplings $t_0$, whereas a central site connects to them through three independently adjustable intracell couplings, $t_1$, $t_2$ and $t_3$. The central site therefore acts as an internal control hub: it introduces the additional coupling channels needed to reconfigure the three-component eigenstates.

When all four coupling scales are balanced ($t_0 = t_1 = t_2 = t_3$), a flat band intersects two linearly dispersing bands at the Brillouin-zone center (Fig. 2b). The resulting threefold crossing forms an isolated pseudospin-1 Dirac point, while the fourth band remains spectrally remote. This separation is important because it allows the low-energy waves to evolve within a well-defined three-state manifold. Redistributing the three intracell couplings lifts the crossing while leaving the remote band detached, reproducing the mass-perturbed spectrum described by equation (1).

The three intracell parameters provide independent control over the magnitude and orientation of the perturbation. While their sum sets the overall energy baseline, their differential ratios dictate the mixing of pseudospin components, thereby rotating the Berry-curvature polarity vector $\mathbf{p}_\Omega$ continuously without rotating the unit cell or the underlying Bravais lattice (Fig. 2c). Cyclically modulating $t_1, t_2, t_3$ rotates $\mathbf{p}_\Omega$ across the full 360° planar range (Fig. 2c).

The spatial mass-gradient $\mathbf{g}_m$ is introduced by gradually modulating the signed amplitude of this coupling perturbation slowly across adjacent cells while preserving the relative ratio that defines $\mathbf{p}_\Omega$. The local mass varies adiabatically from negative to positive across an intermediate zero-mass boundary (Fig. 2d). In this four-site architecture, $\mathbf{p}_\Omega$ is set strictly by the local intracell ratios, whereas $\mathbf{g}_m$ is defined independently by the real-space arrangement of the unit cells. This decoupling enables arbitrary guiding orientations across a fixed physical lattice.

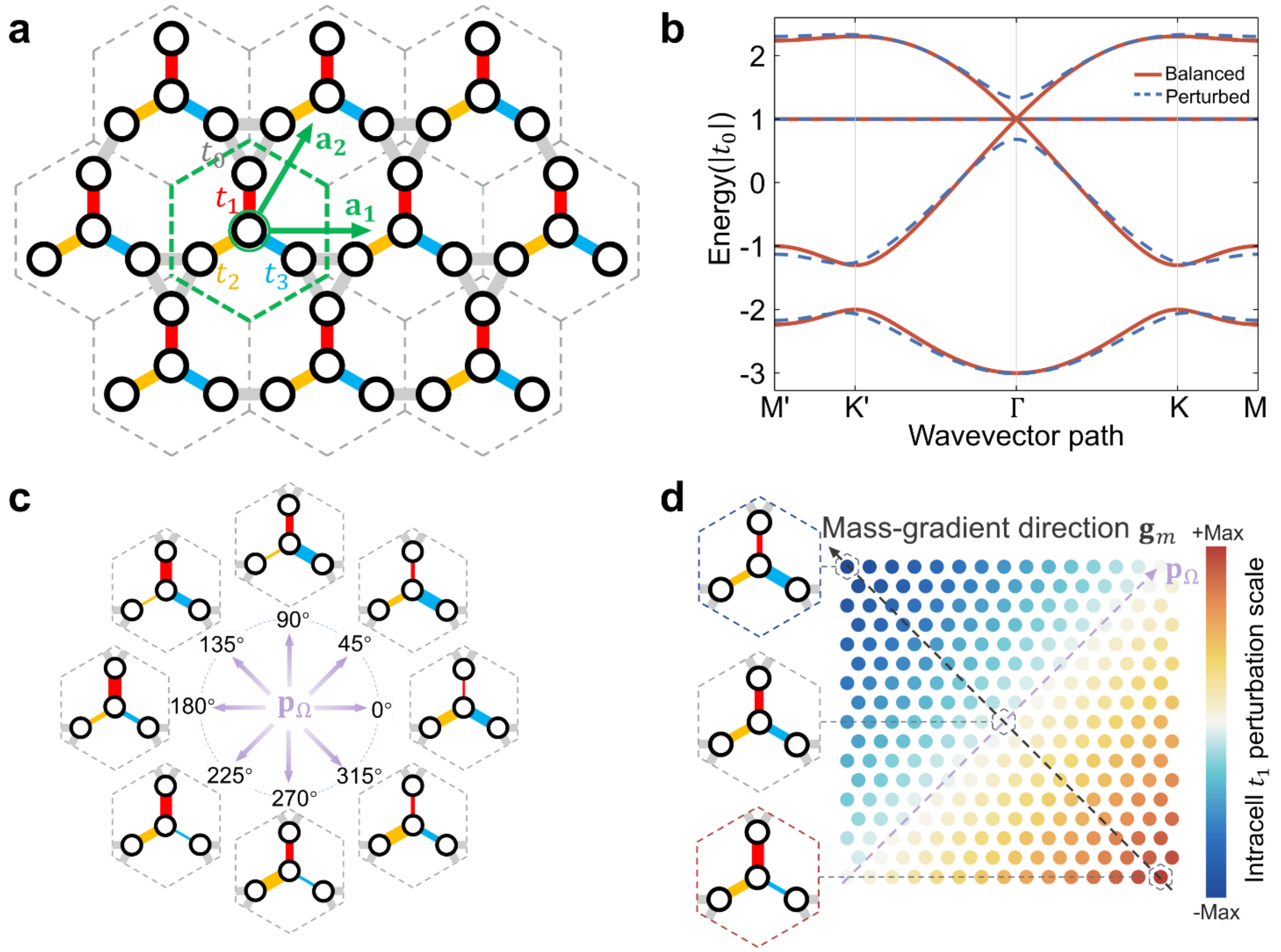

**Fig. 2 The planar lattice for programmable Berry-curvature polarity and adiabatic mass textures. a**, Two-dimensional four-site planar lattice. Fixed intercell links have coupling strength $t_0$, whereas the central site is connected to the three outer sites through independently adjustable intracell couplings $t_1$, $t_2$ and $t_3$, shown in red, yellow and cyan, respectively. The green dashed outline marks a unit cell and $\mathbf{a}_1$ and $\mathbf{a}_2$ are the primitive lattice vectors. **b**, Calculated four-band spectra for balanced couplings (solid red), for which a flat band and two linearly dispersing bands meet at Γ to form an isolated pseudospin-1 Dirac point, and for a representative differential intracell perturbation (dashed blue), which splits the triplet while leaving the fourth band spectrally separated. **c**, Representative intracell-coupling configurations programming Berry-curvature polarity directions $\mathbf{p}_\Omega$ from 0° to 360°. Bond widths encode the relative coupling strengths. The physical orientation of the unit cell remains fixed while the coupling pattern rotates through the plane. **d**, Real-space implementation of an adiabatically graded mass texture. The lattice colours denote the signed perturbation of the intracell coupling $t_1$, from positive (red) to negative (blue), whereas the corresponding mass term evolves in the opposite sense, from negative to positive. Representative unit cells illustrate the associated coupling evolution through the balanced, zero-mass configuration. In this configuration, the Berry-curvature polarity $\mathbf{p}_\Omega$ is oriented at 45°, and $\mathbf{p}_\Omega$ and the mass-gradient direction $\mathbf{g}_m$ are orthogonal with positive orientation, giving the geometric factor $\chi = +1$.

## Experimental observation of reconfigurable pseudospin Hall waveguiding

We experimentally validate this polarity-matching mechanism using CNC-machined acoustic resonator lattices (Fig. 3a). Two complementary lattices were fabricated with identical local unit cells, sharing the same horizontal Berry-curvature polarity $\mathbf{p}_\Omega$, but with reversed mass-gradient direction $\mathbf{g}_m$ (Fig. 3b). This distinction is revealed in the Fourier-resolved acoustic spectra. In the opposite polarity-gradient mismatched configuration ( $\mathbf{p}_\Omega$ and $\mathbf{g}_m$ are orthogonal with negative orientation, corresponding to $\chi = -1$), the mass gradient yields a ladder of transversely localized states, but the two innermost dispersive bands remain separated at $k_x = 0$ (Fig. 3c). Reversing the mass gradient ($\chi = +1$) preserves the higher order bulk bands while driving the lowest pair together, forming a gapless crossing near 9 kHz (Fig. 3d), in close agreement with numerical eigenmodes calculation. The dispersionless feature around the crossing frequency originates from the chiral-symmetry-pinned flat band. The physical signature of the pseudospin Hall response is the deterministic closure of the central gap between the dispersive branches.

Real-space measurements confirm the transport behavior associated with this spectral change. At 9050 Hz, the 0° configuration ($\chi = +1$) exhibits wave propagation horizontally across the lattice while maintaining transverse confinement (Fig. 3e). The field radiates symmetrically away from the source because the phase-symmetric excitation addresses the two reciprocal branches simultaneously. Conversely, in the mismatched configuration ($\chi = -1$), wave transmission into the bulk is suppressed, and acoustic energy remains localized near the source (Supplementary Fig. S8). Bulk embedded waveguiding thus arises specifically from the polarity-enabled linear crossing of the lowest pseudospin branches. To verify directional reconfigurability, we next rotate the Berry-curvature polarity to 45° and construct the corresponding orthogonal mass gradient ($\chi = +1$), keeping the global lattice orientation, source position and external boundary unchanged. As predicted, the measured pressure field realigns along the diagonal direction (Fig. 3f).

Together, the spectral and spatial measurements establish two complementary levels of control. The geometric matching factor $\chi$ dictates whether gapless pseudospin Hall modes are formed, while the orientation of Berry-curvature polarity determines the modes' propagation axis. While the symmetric excitation in Fig. 3 addresses both counterpropagating states simultaneously, tailoring the internal phase structure of the source allows individual pseudospin branches to be launched selectively.

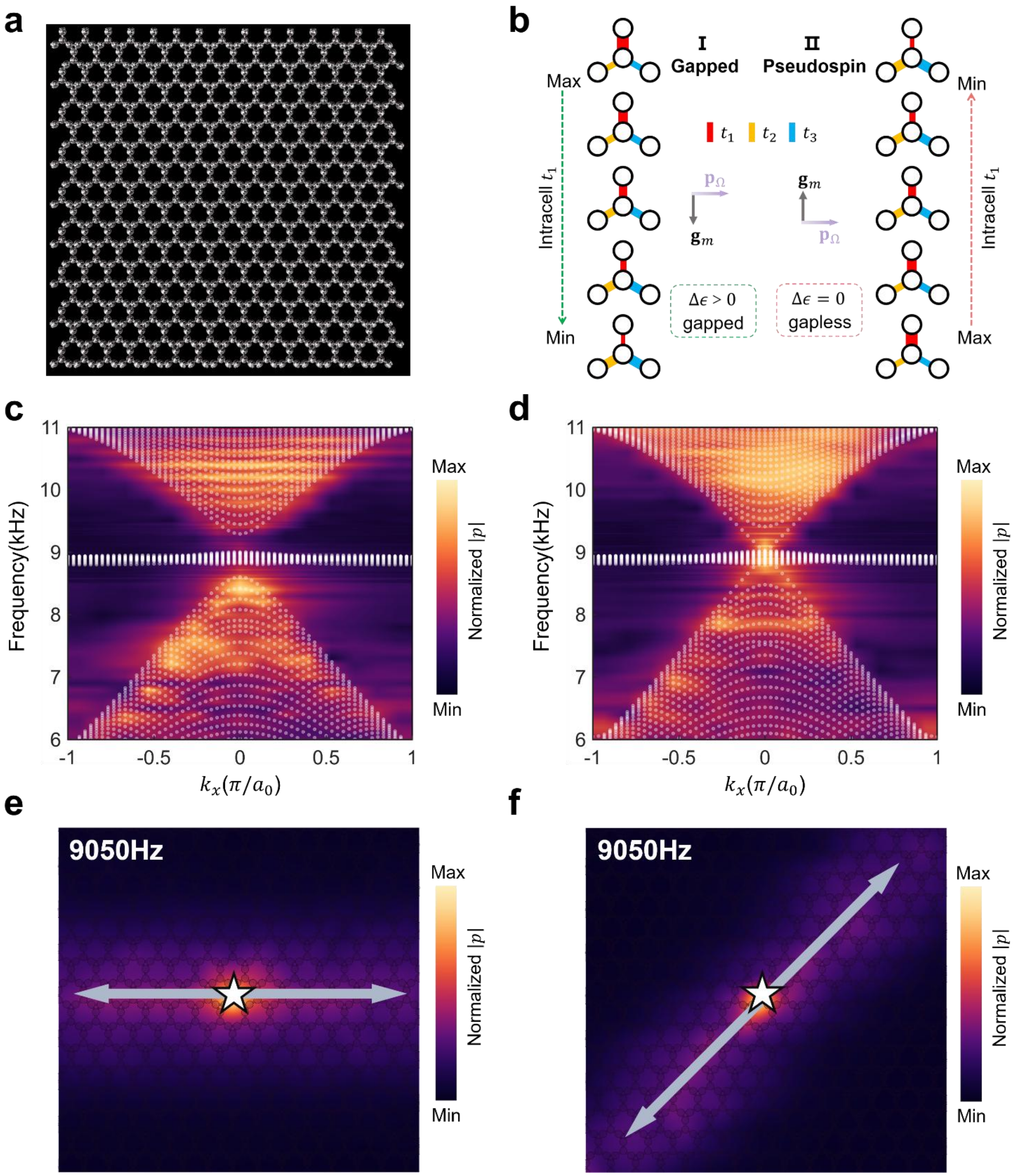

**Fig. 3 Experimental observation of reconfigurable pseudospin Hall waveguiding. a**, Top-view image of the fabricated acoustic lattice implementing spatially graded intracell couplings. **b**, Controlled reversal of the adiabatic mass texture. The two configurations retain the same horizontal Berry-curvature polarity $\mathbf{p}_\Omega$, local mass contrast and confinement strength, whereas reversing the spatial sequence of the unit cells reverses the mass-gradient direction $\mathbf{g}_m$. The mismatched and matched polarity-gradient orientations correspond to the spectra in **c** and **d**, respectively. **c,d,** Measured Fourier-resolved spectra for the mismatched (**c**) and matched (**d**) configurations. Colour denotes the measured spectral intensity and white points show the simulated eigenfrequencies. In **c**,

the two dispersive bands nearest the flat band remain separated at $k_x = 0$; in **d**, the pseudospin bands converge near 9 kHz. Bulk bands are present in both configurations, and the nearly horizontal feature is the flat band. **e,f,** Measured normalized acoustic-pressure amplitudes at 9050 Hz for matched 0° (**e**) and 45° (**f**) designs. Stars mark the excitation position and double-headed arrows indicate the propagation axes. The global lattice orientation and external boundary remain unchanged between (**e**) and (**f**).

## Pseudospin-momentum locking and branch-selective launching

The two branches of the pseudospin Hall pair coexist along the same real-space channel in a reciprocal system, yet carry opposite group velocities and orthogonal internal phase distributions within the pseudospin-1 manifold. A source whose phase profile selectively overlaps with one of these internal mode textures can therefore address an individual propagation branch.

We probe this modal distinction using two mirror-symmetric source configurations, referred to as right- and left-handed excitations. The acoustic lattice, Berry-curvature polarity and mass gradient are held fixed throughout the comparison. For the horizontal waveguide, right-handed excitation concentrates the measured Fourier weight predominantly on the forward-propagation branch, whereas left-handed excitation directs the weight onto the backward propagation branch (Fig. 4a,d). The underlying dispersion curves remain unchanged: the handedness of the source purely controls the projection of power between the two states, confirming that directional transport stems from selective modal overlap.

This selective excitation directly translates into unidirectional propagation in real space. At 9050 Hz, the right-handed excitation launches the horizontal waveguide predominantly towards one side of the source, whereas the left-handed excitation reverses the propagation direction (Fig. 4b,e). The same directional selectivity is observed when the waveguide is reprogrammed to 45°. With the lattice orientation and source geometry held fixed, switching source handedness flips the primary emission between opposing diagonal directions (Fig. 4c,f).

These observations provide direct experimental confirmation of pseudospin-momentum locking: reversing the modal propagation branch switches the internal state to which the source couples, yielding branch-selective excitation. Together with the gap and propagation control established in Fig. 3, this architecture decouples three functions that are usually intertwined in waveguiding: the mass gradient confines the waves, Berry-curvature polarity determines the admissible propagation axis, and the source state selects the counter-propagating branch.

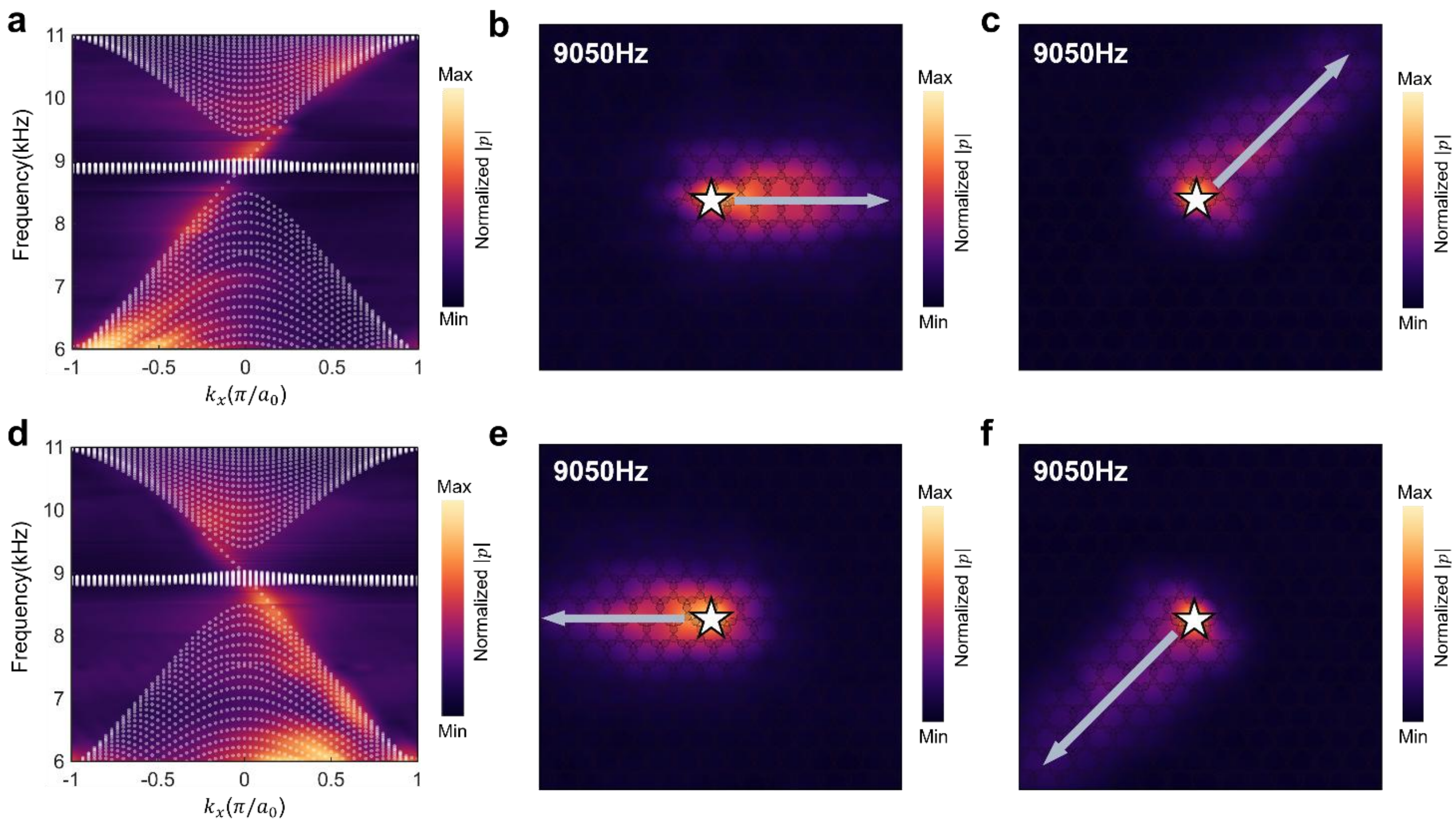

**Fig. 4 Pseudospin-momentum locking and branch-selective launching. a,d,** Measured Fourier-resolved spectra of the 0° pseudospin Hall waveguide under right-handed (**a**) and left-handed (**d**) source-phase configurations. Colour represents measured spectral response and white points denote simulated eigenfrequencies. The complementary excitations concentrate the spectral weight predominantly on the positive- and negative-group velocity branches, respectively. **b,c,** Measured normalized acoustic-pressure amplitudes at 9050 Hz under right-handed excitation for the 0°(**b**) and 45°(**c**) waveguide designs. **e,f,** Corresponding measurements under left-handed excitation for the 0°(**e**) and 45°(**f**) designs, showing reversal of the preferred propagation direction. Stars mark the source position and arrows indicate the dominant transport direction. In the same Berry-curvature polarity configuration, the lattice and adiabatic mass texture are unchanged within each selective-launching comparison; only the relative source phases are varied.

## Discussion

In this work, we have established Berry-curvature polarity as a versatile and programmable degree of freedom governing adiabatic pseudospin Hall transport in two dimensions. Its oriented matching with an adiabatic mass gradient determines whether the dispersive bands open a dispersive gap or drive the system into a gapless, helical pair of pseudospin modes. Because inverting the mass gradient yields qualitatively contrasting spectral and transport responses, this internal geometric property becomes directly accessible through both band dispersion and real-space wave propagation. While the net Berry flux integrated over the Brillouin zone vanishes, the sign-resolved sectors preserve an oriented dipolar polarity. Adiabatic spatial modulation thus converts this local Bloch-state geometry into a deterministic selection rule for robust, bulk-embedded transport.

The four-site lattice architecture makes this selection rule readily engineerable via independent control of the intracell coupling pattern and its spatial arrangement. These controls allow the Berry-curvature polarity and the directed mass variation to be co-designed within a fixed lattice frame, allowing waveguiding paths to be routed arbitrarily without altering the physical boundary of the system.

Furthermore, coupling this reconfigurable propagation axis to helical source phases selectively addresses individual counterpropagating pseudospin branches, locking directional emission to the programmed geometric axis. Channel routing and branch selection thereby operate as independent, complementary controls for tailoring pseudospin-dependent transport across the interior of a planar lattice.

Beyond acoustic metamaterials, the low-energy Hamiltonian and geometric matching rules developed here provides a general framework for multicomponent wave platforms. Photonic, polaritonic, and mechanical lattices designed with a pseudospin-1 kinetic sector and tunable quadrupolar couplings could implement analogous adiabatic mass textures to realize bulk-embedded, reconfigurable waveguides without sharp physical interfaces. More fundamentally, this mechanism demonstrates that waveguiding properties can be programmed through the vector matching of real-space gradients and momentum-space geometric textures, establishing a design principle for topological and non-Abelian wave manipulation across classical and quantum settings.

## Data Availability

All data needed to evaluate the conclusions in the paper are present in the paper and the supplementary material. Source data are available upon reasonable request.

## Author contributions

Qinhui Jiang and Jidong Song contributed to the work equally. All authors contributed to the work substantially.

## Conflict of interest

The authors declare no conflict of interests.

## Acknowledgement

We thank financial support from the National Key R&D Program of China (Grant No. 2023YFB3811400), National Natural Science Foundation of China (Grant No.12404435), the Guangdong Science and Technology Commission (Grant No.2025A1515010774), the New Cornerstone Science Foundation, the Research Grants Council of Hong Kong (AoE/P-502/20, STG3/E-704/23-N, C7015-24G, 17309021), and Guangdong Provincial Quantum Science Strategic Initiative (GDZX2204004, GDZX2304001).